\documentclass[prb,amsmath,amsfonts,amssymb,preprint,letterpaper]{revtex4}
\usepackage{graphicx}
\usepackage{bm}
\usepackage{color}

\begin{document}
\title{Particle-particle and quasiparticle random phase approximations: Connections to coupled cluster theory}
\author{Gustavo E. Scuseria}
\author{Thomas M. Henderson}
\affiliation{Department of Chemistry and Department of Physics and Astronomy, Rice University, Houston, TX 77005-1892}
\author{Ireneusz W. Bulik}
\affiliation{Department of Chemistry, Rice University, Houston, TX 77005-1892}
\date{Updated \today}

\begin{abstract}
We establish a formal connection between the particle-particle (pp) random phase approximation (RPA) and the ladder channel of the coupled cluster doubles (CCD) equations. The relationship between RPA and CCD is best understood within a Bogoliubov quasiparticle (qp) RPA formalism. This work is a follow-up to our previous formal proof on the connection between particle-hole (ph) RPA and ring-CCD. Whereas RPA is a quasibosonic approximation, CC theory is a correct bosonization in the sense that the wavefunction and Hilbert space are exactly fermionic. Coupled cluster theory achieves this goal by interacting the ph (ring) and pp (ladder) diagrams via a third channel that we here call ``crossed-ring'' whose presence allows for full fermionic antisymmetry.  Additionally, coupled cluster incorporates what we call ``mosaic'' terms which can be absorbed into defining a new effective one-body Hamiltonian.  The inclusion of these mosaic terms seems to be quite important.  The pp-RPA and qp-RPA equations are textbook material in nuclear structure physics but are largely unknown in quantum chemistry, where particle number fluctuations and Bogoliubov determinants are rarely used. We believe that the ideas and connections discussed in this paper may help design improved ways of incorporating RPA correlation into density functionals based on a CC perspective.
\end{abstract}
\maketitle

\section{Introduction}
In a previous paper,\cite{Scuseria2009} we established a connection between the particle-hole random phase approximation (RPA) and the ring channel of the coupled cluster doubles (CCD) equations.  Here, we extend this analysis to the ladder channel of CCD and demonstrate a rigorous connection with the particle-particle (pp) RPA equations.  RPA is a quasibosonic approximation in the sense that fermion products are treated as bosons when in reality these operators satisfy Lie algebra commutation rules that are neither fermionic nor bosonic.  The RPA quasibosonic approximation contaminates the fermionic Hilbert space with bosonic states, thus leading to systematic overestimation of ground state correlation energies.  We here also show that from a Bogoliubov quasiparticle (qp) RPA perspective, the particle-hole and particle-particle RPA channels get added together but do not interact.  On the other hand, CCD can be interpreted as a ``correct bosonization'' because a third channel, here referred to as ``crossed ring,'' allows the other two channels to interact and closes the equations in a manner that exactly preserves the fermionic nature of the wavefunction and the Hilbert space of the problem.  The particle-particle RPA and quasiparticle RPA equations are textbook material in nuclear structure physics but are largely unknown in quantum chemistry, where particle number fluctuations and HFB determinants are rarely used.  Coupled cluster theory has undoubtedly been very successful in quantum chemistry and given the current interest of using RPA to improve DFT,\cite{RPAReview1,RPAReview2} we believe that the ideas and connections discussed in this paper may help design improved ways of incorporating RPA effects into functionals based on a coupled cluster perspective.

\section{Different Flavors of RPA}
\subsection{Particle-Hole RPA}
The traditional RPA formulation can be interpreted as an attempt to treat particle-hole fermionic excitations as bosons.  RPA is a theory that can be used both for excited states and ground state correlation.  The standard derivation of the RPA equations usually follows a so-called equation of motion approach.\cite{RingSchuck}  Here, we pursue a somewhat different perspective. Products of particle-hole excitation fermionic operators are assumed to be bosons
\begin{equation}
a_a^\dagger \, a_i \rightarrow b_\beta
\end{equation}
that satisfy commutation rules
\begin{align}
\left[b_\beta, b_{\beta^\prime}^\dagger\right] &= \delta_{\beta \beta^\prime}
\\
\left[b_\beta, b_{\beta^\prime}\right] &= \left[b_\beta^\dagger, b_{\beta^\prime}^\dagger\right] = 0.
\end{align}
Note that these commutation rules are satisfied \textit{on average} when using a single-determinant reference.  

This quasiboson approximation is the central approximation of RPA.  In reality, the $b_{\beta }$ fermionic product operators satisfy unitary group $U(M)$ commutation rules that are those of a Lie algebra
\begin{equation}
\left[a_a^\dagger \, a_i, a_j^\dagger \, a_b \right] = \delta_{ij} \, a_a^\dagger \, a_b - \delta_{ab} \, a_j^\dagger \, a_i.
\end{equation}
We follow the traditional notation where spin orbitals $i,j,k,l$ are occupied (holes) and $a,b,c,d$ are unoccupied (particles) in a reference determinant. Indices $p,q,r,s,\ldots$ refer to unspecified spin orbitals. All repeated indices are summed.

In RPA, the quartic fermionic Hamiltonian
\begin{equation}
H = h_{pq} \, a_p^\dagger \, a_q + \frac{1}{4} \langle pq \| rs \rangle \, a_p^\dagger \, a_q^\dagger \, a_s a_r
\end{equation}
is, after particle-hole transformation, interpreted as a quadratic bosonic Hamiltonian with number-violating terms:
\begin{equation}
H = E_\mathrm{HF} + A_{\beta \beta^\prime} \, b_\beta^\dagger \, b_{\beta^\prime} + \frac{1}{2} \, \left(B^{}_{\beta \beta^\prime} b_\beta^\dagger b_{\beta^\prime}^\dagger + B_{\beta \beta^\prime}^\star \, b_\beta \, b_{\beta^\prime}\right).
\label{bH}
\end{equation}
Here, the Hermitian $\bm{A}=\bm{A}^\dagger$ and symmetric $\bm{B}=\bm{B}^\mathsf{T}$ matrices are those of standard RPA:
\begin{subequations}
\begin{align}
A_{\beta \beta^\prime} &\rightarrow A_{ai,bj} = f_{ab} \, \delta_{ij} - f_{ji} \, \delta_{ab} + \langle aj \| ib \rangle,
\\
B_{\beta \beta^\prime} &\rightarrow B_{ai,bj} = \langle ij \| ab \rangle,
\end{align}
\end{subequations}
where $\bm{f}$ is the Fock matrix.  Using the bosonic commutations relations, one can recognize the Hamiltonian as
\begin{equation}
H = E_\mathrm{HF} + \frac{1}{2} \begin{pmatrix} \bm{b}^\dagger & \bm{b} \end{pmatrix} \begin{pmatrix} \bm{A} & \bm{B} \\ \bm{B}^\star & \bm{A}^\star \end{pmatrix} \begin{pmatrix} \bm{b} \\ \bm{b}^\dagger \end{pmatrix} - \frac{1}{2} \, \mathrm{Tr}(\bm{A}).
\end{equation}
This is a \textit{quadratic} bosonic Hamiltonian that can be solved exactly (diagonalized) via a Bogoliubov transformation for bosons.\cite{BlaizotRipka} The solution is obtained via a non-Hermitian diagonalization problem
\begin{equation}
\begin{pmatrix} \bm{A} & \bm{B} \\ -\bm{B}^\star & -\bm{A}^\star\end{pmatrix}
\begin{pmatrix} \bm{X} & \bm{Y}^\star \\ \bm{Y} &  \bm{X}^\star \end{pmatrix}
=
\begin{pmatrix} \bm{X} & \bm{Y}^\star \\ \bm{Y} &  \bm{X}^\star \end{pmatrix}
\begin{pmatrix} \bm{\omega} & \bm{0} \\ \bm{0} & -\bm{\omega}^\star \end{pmatrix}.
\label{RPA}
\end{equation}
For the solution to be physically meaningful, we must have
\begin{equation}
\bm{M} = \begin{pmatrix} \bm{A} & \bm{B} \\ \bm{B}^\star & \bm{A}^\star\end{pmatrix} >= 0
\end{equation}
so that all of the RPA eigenvalues $\bm{\omega}$ and the correlation energy
\begin{equation}
E_c = \frac{1}{2} \sum_{\omega > 0} \omega - \frac{1}{2} \, \mathrm{Tr}(\bm{A})
\end{equation}
are real.

General quadratic fermionic Hamiltonians (\textit{i.e.} those containing number-conserving and number-violating terms as the bosonic Hamiltonian above) are diagonalized via Bogoliubov canonical transformations defining a Hartree-Fock-Bogoliubov (HFB) determinant, which is a product state of quasiparticles.  The simplest possible bosonic wavefunction is defined by the diagonalization of $\bm{A}$, which is the so-called Hartree-Bose (HB) problem.  Hartree-Bose yields a \textit{condensate} wavefunction where every boson occupies the same orbital so that the density matrix is simply $\rho_{pq} = z_p z_q^\star = (\bm{z} \, \bm{z}^\dagger)_{pq}$.  Including $\bm{B}$ but retaining the condition that every boson occupies the same orbital specified by $\bm{z}$ leads to a coherent state in Fock space, and is the result of a simple shift canonical transformation.  These two wavefunctions are essentially equivalent in the context of RPA.  The full quadratic bosonic Hamiltonian diagonalization involves a Bogoliubov bosonic transformation (shift plus rotation) that introduces fluctuations over the condensate; these fluctuations can be thought of as correlations.  When we approximate fermionic product operators as bosons, we take advantage of this bosonic structure to obtain ground state correlations via a quadratic Hamiltonian diagonalization which yields a \textit{correlated} wavefunction essentially because the bosons are composite fermions.  One disadvantage of this approach is that the dimension of the bosonic Hamiltonian is proportional to the square of the number of fermions (strictly, the number of particle-hole excitations).

From this perspective, the RPA for fermionic excitations and ground-state correlations is a bosonic treatment that includes two fundamental choices:
\begin{itemize}
\item The effective bosonic excitation operators in particle-hole RPA are written as
\begin{equation}
b_\nu^\dagger \rightarrow Q_\nu^\dagger = X_{ai}^{\nu } \, a_a^\dagger \, a_i - Y_{ai}^\nu \, a_i^\dagger \, a_a.
\end{equation}
\item The fermionic reference state in traditional particle-hole RPA is simply the Hartree-Fock determinant, $|\mathrm{HF}\rangle$.
\end{itemize}
Other choices for excitation operators and the reference state are possible.

The fact that products of fermion operators are not bosons manifests in RPA in myriad ways.  To name a few and without going into details, the lack of a killer condition, the difficulty in obtaining a self-consistent ground-state RPA approximation, the indefinition of entire blocks of the two-particle reduced density matrix (RDM), the appearance of non-representable RDMs, violations to Pauli's principle, and the presence of unphysical (bosonic) states in the fermionic Hilbert space of the problem are all symptoms of the same condition: the quasiboson approximation.

\subsection{Particle-Particle RPA}
In this approach, one simply considers non-number--conserving excitations and de-excitations\cite{RingSchuck,BlaizotRipka}
\begin{equation}
Q_\nu^\dagger = \frac{1}{2} X^\nu_{ab} \, a^\dagger_a \, a^\dagger_b - \frac{1}{2} Y^\nu_{ij} a^\dagger_j \, a^\dagger_i.
\end{equation}
As shown below, this leads to a different particle-particle RPA problem, one that is seldom discussed in quantum chemistry.  From a coupled cluster perspective, the contractions in particle-particle RPA are ladders rather than the rings in particle-hole RPA.  A formal analytic proof is also presented showing the equivalence between particle-particle RPA and ladder-CCD, which is a natural follow-up to our previous proof of the equivalence between particle-hole RPA and ring-CCD.\cite{Scuseria2009}  In particle-particle RPA too, the commutation rules between the excitation operators are approximated as being bosonic whereas in reality they are those of an $SO(2M)$ Lie algebra, one that includes the number conserving ph subalgebra of $U(M)$.

While one can cast particle-particle RPA into a symplectic eigenvalue problem, and from it extract a ground-state correlation energy, we do not show the detailed expressions in this subsection.  Rather, we prefer to wait until we have discussed the more general \textit{quasiparticle} RPA, which subsumes both particle-hole and particle-particle RPA into a single diagonalization problem.

\subsection{Quasiparticle RPA}
The particle-hole and particle-particle forms of the excitation operators are special cases of a more general quasiparticle excitation operator that one can write as\cite{RingSchuck}
\begin{equation}
Q_\nu^\dagger = \frac{1}{4} \mathcal{X}_{pq}^\nu \, \alpha_p^\dagger \, \alpha_q^\dagger - \frac{1}{4} \mathcal{Y}_{pq}^\nu \, \alpha_q \, \alpha_p
\label{QPOperator}
\end{equation}
where the $\alpha$ are Bogoliubov canonically transformed fermionic quasiparticle operators
\begin{align}
\begin{pmatrix} \bm{\alpha} \\ \bm{\alpha}^\dagger \end{pmatrix}
&=
\begin{pmatrix} \bm{U}^\dagger & \bm{V}^\dagger \\ \bm{V}^\mathsf{T} & \bm{U}^\mathsf{T} \end{pmatrix}
\begin{pmatrix} \bm{a} \\ \bm{a}^\dagger \end{pmatrix}
=
\bm{W}^\dagger
\begin{pmatrix} \bm{a} \\ \bm{a}^\dagger \end{pmatrix}
\\
\bm{W}^\dagger \, \bm{W} &= \bm{W} \, \bm{W}^\dagger = \bm{1}
\end{align}
preserving anticommutation rules.  In this quasiparticle basis, the Hamiltonian takes the form
\begin{equation}
\begin{split}
H &= H^0
   + H^{11}_{pq} \, \alpha_p^\dagger \, \alpha_q
   + \frac{1}{2} \left(H^{20}_{pq} \, \alpha^\dagger_p \, \alpha^\dagger_q + H_{qp}^{20\star} \, \alpha_q \, \alpha_p\right)
   + \frac{1}{4} \, H^{22}_{pqrs} \, \alpha_p^\dagger \, \alpha_q^\dagger \, \alpha_r \, \alpha_s
\\
  &+ H^{40}_{pqrs} \, \alpha_p^\dagger \, \alpha_q^\dagger \, \alpha_r^\dagger \, \alpha_s^\dagger
   + H^{40\star}_{sqrp} \, \alpha_s \, \alpha_r \, \alpha_q \, \alpha_p
   + H^{31}_{pqrs} \, \alpha_p^\dagger \, \alpha_q^\dagger \, \alpha_r^\dagger \, \alpha_s
   + H^{31\star}_{sqrp} \, \alpha_s^\dagger \, \alpha_r \, \alpha_q \, \alpha_p.
\end{split}
\end{equation}
Detailed expressions for the matrix elements can be found in Appendix E of Ref. \onlinecite{RingSchuck}.  From the quasiparticle mean-field approximation we obtain quasiparticle energies $E_k$.

Assuming a quasiboson approximation for the fermionic quasiparticle products of Eqn. \ref{QPOperator} and taking $|\mathrm{HFB}\rangle$ as a reference, one finds an RPA problem of the form
\begin{equation}
\begin{pmatrix} \bm{\mathcal{A}} & \bm{\mathcal{B}} \\ -\bm{\mathcal{B}}^\star & -\bm{\mathcal{A}}^\star  \end{pmatrix}
\begin{pmatrix} \bm{\mathcal{X}} & \bm{\mathcal{Y}}^\star \\ \bm{\mathcal{Y}} & \bm{\mathcal{X}}^\star \end{pmatrix}
=
\begin{pmatrix} \bm{\mathcal{X}} & \bm{\mathcal{Y}}^\star \\ \bm{\mathcal{Y}} & \bm{\mathcal{X}}^\star \end{pmatrix}
\begin{pmatrix} \bm{\omega} & \bm{0} \\ \bm{0} & -\bm{\omega} \end{pmatrix}
\end{equation}
where the indices now run over all pairs. The RPA matrices are
\begin{subequations}
\begin{align}
\mathcal{A}_{pq,rs} &=  \langle \Big[ \alpha_{q}\alpha_{p} \Big[H,\alpha{}^\dagger_{r}\alpha{}^\dagger_{s}\Big]\Big] \rangle = \left(E_{p}+E_{q}\right) \delta _{pr} \delta_{qs} + H_{pqrs}^{22}
\\
\mathcal{B}_{pq,rs} &= \langle \Big[ \alpha_{q}\alpha_{p} \Big[H,\alpha{}_{s}\alpha{}_{r}\Big]\Big] \rangle =  4! H_{pqrs}^{40}.
\end{align}
\end{subequations}
Again, $\bm{\mathcal{A}}$ is Hermitian and $\bm{\mathcal{B}}$ is symmetric. Note that the quasiparticle RPA matrix is $\bigl(\begin{smallmatrix} \bm{\mathcal{A}} & \bm{\mathcal{B}} \\ -\bm{\mathcal{B}}^\star & -\bm{\mathcal{A}}^\star \end{smallmatrix} \bigr) = \bm{\eta} \, \bm{\mathcal{M}}$, where $\bm{\eta} = \bigl(\begin{smallmatrix} \bm{1} & \bm{0} \\ \bm{0} & \bm{-1} \end{smallmatrix} \bigr)$ is a symplectic metric and $\bm{\mathcal{M}}$ is the HFB orbital Hessian.

The quasiparticle RPA matrix has the standard symplectic form, and therefore has properties similar to particle-hole RPA.  In particular, qp-RPA leads to a Ricatti equation\cite{Scuseria2009}
\begin{equation}
\bm{\mathcal{B}}^\star + \bm{\mathcal{A}}^\star \, \bm{\mathcal{T}} + \bm{\mathcal{T}} \, \bm{\mathcal{A}} + \bm{\mathcal{T}} \, \bm{\mathcal{B}} \, \bm{\mathcal{T}} = \bm{0}
\end{equation}
with
\begin{equation}
\bm{\mathcal{T}} = \bm{\mathcal{Y}} \, \bm{\mathcal{X}}^{-1}
\end{equation}
and a corresponding correlation energy
\begin{equation}
E_c = \frac{1}{2} \sum_{\omega > 0} \omega - \frac{1}{2} \, \mathrm{Tr}(\bm{\mathcal{A}}).
\end{equation}

From the form of $Q_\nu^\dagger$ used in qp-RPA, it is evident that qp-RPA does not treat the $31$ and $13$ blocks of $H$; in other words $H^{31},$ which is responsible for connecting the particle-particle and particle-hole channels, does not appear in qp-RPA.  Below we discuss how single-reference CCD theory makes these channels interact, essentially by demanding that these channels (together with a crossed-ring channel we will introduce later) lead to the same fermionic wave function amplitudes.  We christen this process as a ``correct bosonization.''

Let us now consider a special case of quasiparticle RPA, namely, the limit when the HFB reference determinant reduces to Hartree-Fock.  For systems with purely repulsive electron-electron interactions, this is the variationally optimal result.\cite{LiebHFB,Takashi}  In this case, the RPA matrix greatly simplifies, and takes the blocked form
\begin{equation}
\bm{\eta} \, \bm{\mathcal{M}} = 
\begin{pmatrix}
\bm{\mathcal{A}}_{oo,oo}         &  \bm{0}                          &  \bm{0}                           &
\bm{0}                           &  \bm{0}                          &  \bm{\mathcal{B}}_{oo,vv}        \\
\bm{0}                           &  \bm{\mathcal{A}}_{ov,ov}        &  \bm{0}                           &
\bm{0}                           &  \bm{\mathcal{B}}_{ov,ov}        &  \bm{0}                          \\
\bm{0}                           &  \bm{0}                          &  \bm{\mathcal{A}}_{vv,vv}         &
\bm{\mathcal{B}}_{vv,oo}         &  \bm{0}                          &  \bm{0}                          \\
\bm{0}                           &  \bm{0}                          & -\bm{\mathcal{B}}_{oo,vv}^\star   &
-\bm{\mathcal{A}}_{oo,oo}^\star  &  \bm{0}                          &  \bm{0}                          \\
\bm{0}                           & -\bm{\mathcal{B}}_{ov,ov}^\star  &  \bm{0}                           &
\bm{0}                           & -\bm{\mathcal{A}}_{ov,ov}^\star  &  \bm{0}                          \\
-\bm{\mathcal{B}}_{vv,oo}^\star  &  \bm{0}                          &  \bm{0}                           &
\bm{0}                           &  \bm{0}                          & -\bm{\mathcal{A}}_{vv,vv}^\star
\end{pmatrix}
\end{equation}
where subscripts ``oo'', ``ov'', and ``vv'' refer to occupied-occupied, occupied-virtual, and virtual-virtual, respectively.  The indices here run over unique orbital pairs, so ``oo'' has indices $i < j$, ``ov'' has indices $ia$, and ``vv'' has indices $a < b$.  We should note that when we have a genuine HFB reference which does not conserve particle number, ``occupied'' and ``virtual'' lose their meaning and the indices of the quasiparticle RPA matrix correspond to quasiparticle creation and annihilation operators, as seen from Eqn. \ref{QPOperator}.

To make contact with our later discussion, we note that the matrices $\bm{\mathcal{A}}_{ov,ov}$ and $\bm{\mathcal{B}}_{ov,ov}$ are the matrices $\bm{A}$ and $\bm{B}$ of particle-hole RPA, and we will define
\begin{subequations}
\begin{align}
\bm{\mathcal{A}}_{oo,oo} &= \bm{D},
\\
\bm{\mathcal{A}}_{vv,vv} &= \bm{C},
\\
\bm{\mathcal{B}}_{vv,oo} &= -\bar{\bm{B}}.
\end{align}
\end{subequations}
Using the fact that $\bm{\mathcal{B}}_{oo,vv} = (\bm{\mathcal{B}}_{vv,oo})^\mathsf{T}$, the quasiparticle RPA matrix $\bm{\mathcal{M}}$ expressed in this notation is
\begin{equation}
\bm{\eta} \, \bm{\mathcal{M}} = 
\begin{pmatrix}
\bm{D}              &  \bm{0}        &   \bm{0}                    &
\bm{0}              &  \bm{0}        &  -\bar{\bm{B}}^\mathsf{T}  \\
\bm{0}              &  \bm{A}        &  \bm{0}                     &
\bm{0}              &  \bm{B}        &  \bm{0}                    \\
\bm{0}              &  \bm{0}        &  \bm{C}                     &
-\bar{\bm{B}}       &  \bm{0}        &  \bm{0}                    \\
\bm{0}              &  \bm{0}        &  \bar{\bm{B}}^\dagger       &
-\bm{D}^\star       &  \bm{0}        &  \bm{0}                    \\
\bm{0}              & -\bm{B}^\star  &  \bm{0}                     &
\bm{0}              & -\bm{A}^\star  &  \bm{0}                    \\
\bar{\bm{B}}^\star  &  \bm{0}        &  \bm{0}                     &
\bm{0}              &  \bm{0}        & -\bm{C}^\star
\end{pmatrix}
\end{equation}
The matrix elements of $\bm{C}$, $\bm{D}$, and $\bar{\bm{B}}$ are
\begin{subequations}
\begin{align}
D_{ij,kl} &= -(\epsilon_i + \epsilon_j) \, \delta_{ik} \, \delta_{jl} + \langle kl \| ij \rangle,
\\
C_{ab,cd} &= (\epsilon_a + \epsilon_b) \, \delta_{ac} \, \delta_{bd} + \langle ab \| cd \rangle,
\\
\bar{B}_{ab,ij} &= \langle ab \| ij \rangle.
\end{align}
\end{subequations}

It is clear that one can decompose the quasiparticle RPA into subproblems.  One subproblem gives us the usual particle-hole RPA.  Taking the central block of $\bm{\mathcal{M}}$ gives us particle-particle RPA, wherein one solves\cite{RingSchuck}
\begin{equation}
\begin{pmatrix}\bm{C}  &  -\bar{\bm{B}}  \\ \bar{\bm{B}}^\dagger & -\bm{D}^\star \end{pmatrix}
\begin{pmatrix}\bm{X}_1 & \bm{Y}_2 \\ \bm{Y}_1 & \bm{X}_2 \end{pmatrix}
=
\begin{pmatrix}\bm{X}_1 & \bm{Y}_2 \\ \bm{Y}_1 & \bm{X}_2 \end{pmatrix}
\begin{pmatrix}\bm{\Omega_1} & \bm{0} \\ \bm{0} & \bm{\Omega}_2 \end{pmatrix}.
\end{equation}
The frequencies $\bm{\Omega}_1$ are positive and $\bm{\Omega}_2$ are negative.  The remaining portion of $\bm{\mathcal{M}}$ gives hole-hole RPA, the symplectic counterpart to particle-particle RPA.  Note that particle-particle and hole-hole RPA are not \textit{individually} symplectic eigenvalue problems, so one cannot straightforwardly extract a plasmonic correlation energy from just one or the other.  Rather, they should be grouped together as
\begin{equation}
\begin{pmatrix}
\bm{D}              & \bm{0}                &  \bm{0}          &  -\bar{\bm{B}}^\mathsf{T}   \\
\bm{0}              & \bm{C}                &  -\bar{\bm{B}}   &  \bm{0}                     \\
\bm{0}              & \bar{\bm{B}}^\dagger  & -\bm{D}^\star    &  \bm{0}                     \\
\bar{\bm{B}}^\star  & \bm{0}                &  \bm{0}          &  -\bm{C}^\star
\end{pmatrix}
\begin{pmatrix}
\bm{X}_2^\star  &  \bm{0}    &  \bm{0}    &  \bm{Y}_1^\star  \\
\bm{0}          &  \bm{X}_1  &  \bm{Y}_2  &  \bm{0}          \\
\bm{0}          &  \bm{Y}_1  &  \bm{X}_2  &  \bm{0}          \\
\bm{Y}_2^\star  &  \bm{0}    &  \bm{0}    &  \bm{X}_1^\star
\end{pmatrix}
=
\begin{pmatrix}
\bm{X}_2^\star  &  \bm{0}    &  \bm{0}    &  \bm{Y}_1^\star  \\
\bm{0}          &  \bm{X}_1  &  \bm{Y}_2  &  \bm{0}          \\
\bm{0}          &  \bm{Y}_1  &  \bm{X}_2  &  \bm{0}          \\
\bm{Y}_2^\star  &  \bm{0}    &  \bm{0}    &  \bm{X}_1^\star
\end{pmatrix}
\begin{pmatrix}
-\bm{\Omega}_2  &  \bm{0}         &  \bm{0}         &  \bm{0}  \\
\bm{0}          &  \bm{\Omega}_1  &  \bm{0}         &  \bm{0}  \\
\bm{0}          &  \bm{0}         &  \bm{\Omega}_2  &  \bm{0}  \\
\bm{0}          &  \bm{0}         &  \bm{0}         & -\bm{\Omega}_1
\end{pmatrix}.
\end{equation}
From here, the plasmonic correlation energy is just
\begin{equation}
E_c = \frac{1}{2} \mathrm{Tr}\left(\bm{\Omega}_1 - \bm{\Omega}_2 - \bm{C} - \bm{D}\right).
\end{equation}
Using the particle-particle RPA equations, one sees that
\begin{equation}
\mathrm{Tr}(\bm{C}) - \mathrm{Tr}(\bm{D}^\star) = \mathrm{Tr}(\bm{\Omega}_1) + \mathrm{Tr}(\bm{\Omega}_2)
\end{equation}
so that
\begin{equation}
\mathrm{Tr}(\bm{\Omega}_1 - \bm{C}) = -\mathrm{Tr}(\bm{\Omega}_2 + \bm{D}^\star).
\end{equation}
Since $\bm{D}$ is Hermitian so that $\mathrm{Tr}(\bm{D}) = \mathrm{Tr}(\bm{D}^\star)$, it follows that the plasmonic correlation energy associated with particle-particle/hole-hole RPA can be equivalently expressed as
\begin{equation}
E_c = \mathrm{Tr}(\bm{\Omega}_1 - \bm{C}) = -\mathrm{Tr}(\bm{\Omega}_2 + \bm{D}^\star).
\end{equation}

Note finally that because quasiparticle RPA in this limit can be factored into two symplectic subproblems, the correlation energy associated with quasiparticle RPA is simply additive:
\begin{equation}
E_{c,qp-RPA} = E_{c,ph-RPA} + E_{c,pp-RPA}.
\end{equation}

\subsection{Stability of RPA Problems}
The three types of RPA we have discussed above all have the same formal symplectic structure when the particle-particle RPA is understood as particle-particle/hole-hole RPA.  All diagonalize a matrix 
\begin{equation}
\bm{\eta} \, \mathbf{M} = \begin{pmatrix} \mathbf{A} & \mathbf{B} \\ -\mathbf{B}^\star & -\mathbf{A}^\star \end{pmatrix}
\end{equation}
where we recall that
\begin{equation}
\bm{\eta} = \begin{pmatrix} \bm{1} & \bm{0} \\ \bm{0} & -\bm{1}\end{pmatrix}.
\end{equation}
The matrix $\mathbf{M}$ is an orbital Hessian, or, to put it another way, it is a stability matrix.  For particle-hole RPA, diagonalizing $\mathbf{M}$ tests for instabilities of the reference $|\mathrm{HF}\rangle$ solution toward another Hartree-Fock state.  For particle-particle/hole-hole RPA, diagonalizing $\mathbf{M}$ tests for instabilities of the reference $|\mathrm{HF}\rangle$ towards an HFB state, while for quasiparticle RPA, diagonalizing $\mathbf{M}$ tests for instabilities of the reference $|\mathrm{HFB}\rangle$ toward another HFB state.  Stability in this context means that $\mathbf{M} >= 0$.  Earlier, we noted that particle-hole RPA gives physically meaningful results when this condition is satisfied; the same holds for particle-particle and for quasiparticle RPA.  Quite generally, when $\mathbf{M}$ is positive definite, the corresponding RPA problem has real eigenvalues and $2n$ linearly independent eigenvectors where $\mathbf{M}$ is $2n \times 2n$.\cite{Colpa}

From a minor modification of the proof in Appendix 5 of Ref. \onlinecite{Scuseria2009}, we see that a Riccati equation of the form $\mathbf{B}^\star + \mathbf{A}^\star \, \mathbf{T} + \mathbf{T} \, \mathbf{A} + \mathbf{T} \, \mathbf{B} \, \mathbf{T} = \mathbf{0}$ implies an eigenvalue-like problem
\begin{equation}
\begin{pmatrix} \mathbf{A} & \mathbf{B} \\ -\mathbf{B}^\star & -\mathbf{A}^\star \end{pmatrix}
\begin{pmatrix} \mathbf{x} \\ \mathbf{y} \end{pmatrix}
=
\begin{pmatrix} \mathbf{x} \\ \mathbf{y} \end{pmatrix}
\bm{\Delta}.
\end{equation}
Here, we have written
\begin{subequations}
\begin{align}
\mathbf{A} + \mathbf{B} \, \mathbf{T} &= \mathbf{x} \, \bm{\Delta} \, \mathbf{x}^\dagger
\\
\mathbf{y} &= \mathbf{T} \, \mathbf{x}
\end{align}
\end{subequations}
where $\mathbf{x}$ is unitary and $\bm{\Delta}$ is upper triangular.  When the matrix $\bm{\eta} \mathbf{M}$ is diagonalizable, the Riccati equation implies the eigenvalue problem itself.  Whether $\bm{\eta} \mathbf{M}$ is diagonalizable or not, we have
\begin{equation}
\mathrm{Tr}(\mathbf{B} \, \mathbf{T}) = \mathrm{Tr}(\bm{\Delta} - \mathbf{A})
\end{equation}
where we recall that the diagonal elements of $\bm{\Delta}$ are its eigenvalues.

In other words, when the RPA matrix $\bm{\eta} \mathbf{M}$ is diagonalizable, the eigenvectors of the RPA problem are intimately connected to a corresponding Riccati equation.  When the reference is stable, the RPA matrix is diagonalizable.  We cannot say much about the stability of a general Hartree-Fock or HFB determinant, but we note that for a repulsive two-body interaction, HF is never unstable toward HFB.\cite{LiebHFB,Takashi}  Thus, particle-particle RPA applies to the standard Coulombic Hamiltonian should not suffer from complex correlation energies.  A more detailed proof is forthcoming.  In the following sections, we will use a simpler proof which assumes invertibility of $\mathbf{X}$ to derive the Riccati equation from the RPA eigenvalue problem, but the foregoing shows that we do not need $\mathbf{X}^{-1}$ to exist for a relation between RPA and a Riccati equation to be found.

\section{Different Flavors of CCD}
Having discussed RPA at some length, we now turn to coupled cluster doubles theory.\cite{BartlettShavitt}  As we shall see, CCD contains various pieces of the RPA problem and unifies them all in such a way as to preserve the fermionic character of the wavefunction.  It will be convenient for our purposes to work with the Brueckner version of CCD theory (referred to as BD) that eliminates single excitations. This section closely follows the notation of Ref. \onlinecite{GES-IJQC95}. The basic ingredients of the BD model are one ($h$) and two-electron ($v$) integrals, and cluster amplitudes ($t$) in the spin-orbital basis
\begin{subequations}
\begin{align}
f_p^q &= h_p^q + v_{pq}^{qk},
\\
v^{ij}_{ab} &= \langle ij \| ab \rangle = \langle ij | ab \rangle - \langle ij | ba \rangle,
\\
v^{ab}_{ij} &= \langle ab \| ij \rangle = \langle ij \| ab \rangle^\star = \left(v^{ij}_{ab}\right)^\star,
\\
t_{ij}^{ab} &= \langle ab | t_2 | ij \rangle = -t_{ji}^{ab} = -t_{ij}^{ba} = t_{ji}^{ba},
\\
E &= E_0 + E_c,
\\
E_0 &= \frac{1}{2} \, \left(h^i_i + f^i_i\right) = h^i_i + \frac{1}{2} v^{ik}_{ik},
\\
E_c &= \frac{1}{4} \, v^{ij}_{ab} \, t_{ij}^{ab}.
\label{ECCD}
\end{align}
\end{subequations}
Repeated indices are always summed (even in $h_i^i$).  \textit{Upper} and \textit{lower} indices can be identified as \textit{bra} and \textit{ket}, respectively.

We now define a Brueckner effective one-body Hamiltonian through the energy expression
\begin{equation}
E = \frac{1}{2} \left(h_i^i + F_i^i\right)
\end{equation}
which defines the occupied-occupied block of $\bm{F}$
\begin{equation}
F_i^k=f_i^k + \frac{1}{2} \, v_{ab}^{kj} \, t_{ij}^{ab}
\label{FOO}
\end{equation}
Using particle-hole symmetry, we impose
\begin{equation}
F_c^a=f_c^a - \frac{1}{2} \, v_{cb}^{ij} \, t_{ij}^{ab}
\label{FVV}
\end{equation}
which is needed to complete the CCD equations in the desired form (\textit{vide infra}). The occupied-virtual block of $F$, which we force to be zero, is chosen as the $T_1$ equation within the BD approximation (\textit{i.e.}, enforcing zero $T_1$ amplitudes)
\begin{equation}
F_i^a = f_i^a + f_b^j \, t_{ij}^{ab} + \frac{1}{2} \, v_{bc}^{aj} \, t_{ij}^{bc} - \frac{1}{2} \, v_{ib}^{jk} \, t_{jk}^{ab} = 0.
\label{FOV}
\end{equation}
The three-index contraction terms (e.g., $v_{ab}^{kj} \, t_{ij}^{ab}$) appearing in the $F$ equations above are here referred to as \textit{mosaic} (they have ring and ladder contractions). The BD equations become simply
\begin{equation}
\begin{split}
0 &= v_{ij}^{ab}
   + \frac{1}{2} \, t_{kl}^{ab} \, v_{ij}^{kl}
   + \frac{1}{2} \, v_{cd}^{ab} \, t_{ij}^{cd}
   + \frac{1}{4} \, t_{kl}^{ab} \, v_{cd}^{kl} \, t_{ij}^{cd}
\\
  &- P_{ij}\left(t_{kj}^{ab} \, F_i^k\right)
   + P_{ab}\left(F_c^a \, y_{ij}^{cb}\right)
   + P_{ij} P_{ab}\left[\left(v_{ic}^{ak} + \frac{1}{2} \, t_{il}^{ad} \, v_{cd}^{kl}\right) t_{kj}^{cb}\right]
\end{split}
\end{equation}
where $P$ is an index permutation operator (e.g., $P_{ij}=1 -i\leftrightarrow j$).

Let us now analyze the BD amplitude equation:
\begin{itemize}
\item The first term $v_{ij}^{ab}$ is called the \textit{driver}.
\item The next three terms contain pp or hh (\textit{ladder}) contractions only.
\item The third term is quadratic in the amplitudes; this is the highest degree of the equations.
\item The next two terms are most readily understood in the Brueckner canonical basis where $F$ is diagonal with eigenvalues $\zeta$:
\begin{equation}
P_{ab} \left(F^a_c \, t_{ij}^{cb}\right)
 = F^a_c \, t_{ij}^{cb} - F^b_c \, t_{ij}^{ca}
 = F^a_c \, t_{ij}^{cb} + F^b_c \, t_{ij}^{ac}
 = \left(\zeta_a + \zeta_b\right) t_{ij}^{ab}
\end{equation}
and
\begin{equation}
-P_{ij} \left(F_i^k \, t_{kj}^{ab}\right) =-\left(\zeta _i + \zeta_j\right) t_{ij}^{ab}.
\end{equation}
One should note that in these terms, the indices on the eigenvalues $\zeta$ are not summed.  These terms provide the denominators in perturbation theory so they are normally grouped with the \textit{driver} terms.  We use $\zeta$ for the eigenvalues of $F$ to emphasize that these eigenvalues are not the eigenvalues of the Fock operator, and contain a dependence on the $t$ amplitudes.
\item In the double permutation, we find all the \textit{ring} (ph) contractions.  There are eight terms in total:
\begin{equation}
\begin{split}
P_{ij} P_{ab}\left[\left(v_{ic}^{ak} + \frac{1}{2} \, t_{il}^{ad} \, v_{cd}^{kl}\right) t_{kj}^{cb}\right]
 &= v_{ic}^{ak} \, t_{kj}^{cb} + \frac{1}{2} \, t_{il}^{ad} \, v_{cd}^{kl} \, t_{kj}^{cb}
  + v_{jc}^{bk} \, t_{ki}^{ca} + \frac{1}{2} \, t_{jl}^{bd} \, v_{cd}^{kl} \, t_{ki}^{ca}
\\
  &- v_{ic}^{bk} \, t_{kj}^{ca} - \frac{1}{2} \, t_{il}^{bd} \, v_{cd}^{kl} \, t_{kj}^{ca}
   - v_{jc}^{ak} \, t_{ki}^{cb} - \frac{1}{2} \, t_{jl}^{ad} \, v_{cd}^{kl} \, t_{ki}^{cb}.
\end{split}
\end{equation}
The two quadratic terms in the top row are identical as are the two quadratic terms in the bottom row, and using antisymmetry of $v$ and $t$, we can simplify this slightly to
\begin{equation}
\begin{split}
P_{ij} P_{ab}\left[\left(v_{ic}^{ak} + \frac{1}{2} \, t_{il}^{ad} \, v_{cd}^{kl}\right) t_{kj}^{cb}\right]
 &= v_{ic}^{ak} \, t_{kj}^{cb} + v_{jc}^{bk} \, t_{ki}^{ca} + t_{ki}^{ca} \, v_{cd}^{kl} \, t_{lj}^{db}
\\
  &- v_{ic}^{kb} \, t_{kj}^{ac} - v_{jc}^{ka} \, t_{ki}^{bc} + t_{ki}^{bc} \, v_{dc}^{kl} \, t_{lj}^{ad}.
\end{split}
\end{equation}
The terms in the first row are \textit{ring} terms and are included in particle-hole RPA; those in the second row also involve particle-hole contractions, but with summation (and external) indices crossed, so we refer to these as \textit{crossed ring} terms.  Note that including the ring terms but excluding the crossed rings, thereby including only a portion of the antisymmetry term $P_{ij} P_{ab} [\ldots]$, breaks the antisymmetry of the amplitude equations and therefore of the $T_2$ amplitudes.
\end{itemize}

\subsection{Ring-CCD}
Here, we merely summarize the results of Ref. \onlinecite{Scuseria2009}.  We collect the driving term and the ring terms (but not the crossed ring terms) in the ring-CCD equation:
\begin{subequations}
\begin{align}
0 &= v_{ij}^{ab} - P_{ij} \left(F_i^k \, t_{kj}^{ab}\right) + P_{ab} \left(F^a_c \, t_{ij}^{cb}\right) + v_{ic}^{ak} \, t_{kj}^{cb} + v_{cj}^{kb} \, t_{ik}^{ac} + t_{ik}^{ac} \, v^{kl}_{cd} \, t_{lj}^{db}
\\
  &= v_{ij}^{ab} + \left(F^a_c \, \delta_i^k - F_i^k \, \delta^a_c\right) t_{kj}^{cb}
               + \left(F^b_c \, \delta_j^k - F_j^k \, \delta^b_c\right) t_{ik}^{ac}
               + v_{ic}^{ak} \, t_{kj}^{cb} + v_{kj}^{cb} \, t_{ik}^{ac} + t_{ik}^{ac} \, v^{kl}_{cd} \, t_{lj}^{db}.
\end{align}
\end{subequations}
The resulting $T_2$ amplitudes are not antisymmetric but do retain the \textit{bosonic} symmetry $t_{ij}^{ab} = t_{ji}^{ba}$.  We can simplify the amplitudes equations by using the $\bm{A}$ and $\bm{B}$ matrices of particle-hole RPA, which in this notation are
\begin{subequations}
\begin{align}
A_{ia,jb} &= F^b_a \, \delta^i_j - F^i_j \, \delta^b_a + v^{ib}_{aj},
\\
B_{ia,jb} &= v^{ij}_{ab}.
\end{align}
\end{subequations}
Note that we have used the Brueckner Hamiltonian $\bm{F}$ in defining these matrices, though one can instead use the Hartree-Fock Hamiltonian $\bm{f}$.  It is apparent that the ring-CCD equation can be written as
\begin{equation}
\bm{B}^\star + \bm{A}^\star \, \bm{T} + \bm{T} \, \bm{A} + \bm{T} \, \bm{B} \, \bm{T} = \bm{0};
\end{equation}
replacing the Brueckner Hamiltonian with the Hartree-Fock Hamiltonian corresponds to discarding the mosaic terms in the ring-CCD equations.

On the other hand, the particle-hole RPA equations for the non-negative excitation energies are
\begin{equation}
\begin{pmatrix} \bm{A} & \bm{B} \\ -\bm{B}^\star & -\bm{A}^\star \end{pmatrix}
\begin{pmatrix} \bm{X} \\ \bm{Y} \end{pmatrix}
=
\begin{pmatrix} \bm{X} \\ \bm{Y} \end{pmatrix}
\bm{\omega}.
\end{equation}
We have assumed that we are in a physically meaningful case where $\bm{\omega}$ is real, so that we can always choose non-negative $\bm{\omega}$.  We emphasize again that this RPA is a \textit{bosonic} mean-field problem where the Hermitian $\bm{A}$ and symmetric $\bm{B}$ play the role of Fock and pairing fields, respectively.

The equivalence between ring-CCD and particle-hole RPA is most simply established when $\bm{X}$ is invertible, so that we can define $\bm{T} = \bm{Y} \, \bm{X}^{-1}$.  In that case, the RPA eigenvalue problem can be used to derive
\begin{subequations}
\begin{align}
\bm{A} + \bm{B} \, \bm{T} &= \bm{X} \, \bm{\omega} \, \bm{X}^{-1} = \bm{R},
\label{phRPA1}
\\
\bm{B}^\star + \bm{A}^\star \, \bm{T} &= -\bm{T} \, \bm{R}.
\end{align}
\end{subequations}
Inserting the first equation into the second and rearranging yields
\begin{equation}
\bm{B}^\star + \bm{A}^\star \, \bm{T} + \bm{T} \, \bm{A} + \bm{T} \, \bm{B} \, \bm{T} = \bm{0}
\end{equation}
so that from particle-hole RPA we can extract the amplitudes $\bm{T}$ which solve the ring-CCD equation.  Moreover, the RPA correlation energy comes from the plasmon formula,
\begin{equation}
E_c^\mathrm{RPA} = \frac{1}{2} \, \mathrm{Tr}(\bm{\omega} - \bm{A})
\end{equation}
while the coupled cluster correlation energy is just
\begin{equation}
E_c^\mathrm{CCD} = \frac{1}{4} \, t_{ij}^{ab} \, v^{ij}_{ab} = \frac{1}{4} \, \mathrm{Tr}(\bm{B} \, \bm{T}).
\end{equation}
From Eqn. \ref{phRPA1}, we see that
\begin{equation}
\mathrm{Tr}(\bm{\omega} - \bm{A}) = \mathrm{Tr}(\bm{B} \, \bm{T}).
\end{equation}
Thus, the ring-CCD and particle-hole RPA correlation energies differ by a factor of two.  This reflects the fact that ring CCD is not a correct bosonization of the fermionic problem.  The discrepancy in the correlation energy disappears for \textit{direct} RPA, where we keep only Hartree terms in the interaction.

\subsection{Ladder-CCD}
We have seen that the ring-CCD problem is intimately connected to particle-hole RPA; here, we demonstrate that the ladder-CCD problem is analogously connected to particle-particle RPA.   To the best of our knowledge, this connection has never been discussed in the literature.

The ladder-CCD equations are
\begin{subequations}
\begin{align}
0 &= v_{ij}^{ab} - P_{ij} \left(F^k_i \, t_{kj}^{ab}\right) + P_{ab} \left(F^a_c \, t_{ij}^{cb}\right) + \frac{1}{2} \, t_{kl}^{ab} \, v^{kl}_{ij} + \frac{1}{2} \, v^{ab}_{cd} \, t_{ij}^{cd} + \frac{1}{4} \, t_{ij}^{cd} \, v_{cd}^{kl} \, t_{kl}^{ab}
\\
  &= v_{ij}^{ab} + \left(F^a_c \, \delta^b_d + F^b_d \, \delta^a_c\right) t_{ij}^{cd} - \left(F^k_i \, \delta^l_j + F^l_j \, \delta^k_i\right) \, t_{kl}^{ab} + \frac{1}{2} \, t_{kl}^{ab} \, v^{kl}_{ij} + \frac{1}{2} \, v^{ab}_{cd} \, t_{ij}^{cd} + \frac{1}{4} \, t_{ij}^{cd} \, v_{cd}^{kl} \, t_{kl}^{ab}.
\end{align}
\end{subequations}
We can express this in terms of the matrices $\bar{\bm{B}}$, $\bm{C}$, and $\bm{D}$ of particle-particle RPA, which in this notation are
\begin{subequations}
\begin{align}
\bar{B}_{ab,ij} &= v^{ab}_{ij}
\\
C_{ab,cd} &= \left(F^a_c \, \delta^b_d + F^b_d \, \delta^a_c\right) + v^{ab}_{cd},
\\
D_{ij,kl} &= -\left(F^k_i \, \delta^l_j + F^l_j \, \delta^k_i\right) + v^{kl}_{ij},
\end{align}
\end{subequations}
where we recall that $\bar{\bm{B}}$ is rectangular and $\bm{C}$ and $\bm{D}$ are Hermitian.  Note that while $\bar{\bm{B}}$ and $\bm{B}$ have the same matrix elements, they are organized into bosonic composite indices differently.

With these definitions in hand, the ladder-CCD equations become
\begin{equation}
\bar{\bm{B}} + \bm{C} \, \bm{T} + \bm{T} \, \bm{D}^\star + \bm{T} \, \bar{\bm{B}}^\dagger \, \bm{T} = \bm{0},
\end{equation}
where the composite indices restrict $a < b$ and $i < j$, yielding the needed factors of $\tfrac{1}{2}$ and $\tfrac{1}{4}$.  Here, we have clearly defined $\bm{T}$ as being a $vv \times oo$ matrix.  Equivalently, we could have written
\begin{equation}
\bar{\bm{B}}^\dagger + \tilde{\bm{T}} \, \bm{C} + \bm{D}^\star \, \tilde{\bm{T}} + \tilde{\bm{T}} \, \bar{\bm{B}} \, \tilde{\bm{T}} = \bm{0},
\end{equation}
where $\tilde{\bm{T}} = \bm{T}^\dagger$ is $oo \times vv$.

Recall that the particle-particle RPA problem is
\begin{equation}
\begin{pmatrix}\bm{C}  &  -\bar{\bm{B}}  \\ \bar{\bm{B}}^\dagger & -\bm{D}^\star \end{pmatrix}
\begin{pmatrix}\bm{X}_1 & \bm{Y}_2 \\ \bm{Y}_1 & \bm{X}_2 \end{pmatrix}
=
\begin{pmatrix}\bm{X}_1 & \bm{Y}_2 \\ \bm{Y}_1 & \bm{X}_2 \end{pmatrix}
\begin{pmatrix}\bm{\Omega_1} & \bm{0} \\ \bm{0} & \bm{\Omega}_2 \end{pmatrix},
\end{equation}
and that $\bm{\Omega}_1$ is positive while $\bm{\Omega}_2$ is negative.  From particle-particle RPA, we can write two Riccati equations.  One writes $\bm{T}_1 = -\bm{Y}_1 \, \bm{X}_1^{-1}$ and the other $\bm{T}_2 = -\bm{Y}_2 \, \bm{X}_2^{-1}$; these matrices $\bm{T}_1$ and $\bm{T}_2$ are of dimension $vv \times oo$ and $oo \times vv$, repectively.

The Riccati equation for $\bm{T}_1$ follows from the particle-particle RPA problem for $\bm{X}_1$ and $\bm{Y}_1$, which yields
\begin{subequations}
\begin{align}
\bm{C} + \bar{\bm{B}} \, \bm{T}_1 &= \bm{X}_1 \, \bm{\Omega}_1 \, \bm{X}_1^{-1} = \bm{R}_1,
\\
\bar{\bm{B}}^\dagger + \bm{D}^\star \, \bm{T}_1 &= -\bm{T}_1 \, \bm{R}_1,
\end{align}
\end{subequations}
from which one extracts
\begin{subequations}
\begin{align}
\bm{0} &= \bar{\bm{B}}^\dagger + \bm{D}^\star \, \bm{T}_1 + \bm{T}_1 \, \bm{C} + \bm{T}_1 \, \bar{\bm{B}} \, \bm{T}_1,
\\
\mathrm{Tr}(\bar{\bm{B}} \, \bm{T}_1) &= \mathrm{Tr}(\bm{\Omega}_1 - \bm{C}).
\end{align}
\end{subequations}
If we instead use the particle-particle RPA problem for $\bm{X}_2$ and $\bm{Y}_2$, we get
\begin{subequations}
\begin{align}
-\bar{\bm{B}}^\dagger \, \bm{T}_2 - \bm{D}^\star &= \bm{X}_2 \, \bm{\Omega}_2 \, \bm{X}_2^{-1} = \bm{R}_2,
\\
-\bm{C} \, \bm{T}_2 - \bar{\bm{B}} &= -\bm{T}_2 \, \bm{R}_2
\end{align}
\end{subequations}
which imply that
\begin{subequations}
\begin{align}
\bm{0} &= \bar{\bm{B}} + \bm{C} \, \bm{T}_2 + \bm{T}_2 \, \bm{D}^\star + \bm{T}_2 \, \bar{\bm{B}}^\dagger \, \bm{T}_2,
\\
\mathrm{Tr}(\bar{\bm{B}}^\dagger \, \bm{T}_2) &= -\mathrm{Tr}(\bm{\Omega}_2 + \bm{D}^\star).
\end{align}
\end{subequations}
Clearly, $\bm{T}_2 = \bm{T}_1^\dagger$.

Finally, the ladder-CCD correlation energy is
\begin{equation}
\begin{split}
E_c &= \mathrm{Tr}(\bar{\bm{B}} \, \bm{T}^\dagger) = \mathrm{Tr}(\bar{\bm{B}}^\dagger \, \bm{T}) = \mathrm{Tr}(\bm{\Omega}_1 - \bm{C}) = -\mathrm{Tr}(\bm{\Omega}_2 + \bm{D}^\star)
\\
    &= \frac{1}{2} \mathrm{Tr}\left(\bm{\Omega}_1 - \bm{\Omega}_2 - \bm{C} - \bm{D^\star}\right)
\end{split}
\end{equation}
which is exactly the result from particle-particle RPA.  The formal equivalence proven here is validated numerically by the calculations we discuss later.

\subsection{A Third Channel: Crossed-Ring--CCD}
Comparing the ring- and ladder-CCD equations to the full CCD equations reveals that we have used the driving terms twice, and have excluded the crossed ring terms entirely.  We can take these remaining pieces and form a third sort of channel.  With this channel we can associate a Riccati equation and form an RPA-like problem; this RPA-like problem, however, has no clear physical significance (unlike particle-hole and particle-particle RPA).

Because we have counted the driving terms twice, we would like the crossed-ring--CCD to take the driving term with a minus sign so that the CCD equations for the three channels add to give the regular CCD equations; again, all of this is subject to caveats with regards to the mosaic terms, which should be included in decomposing the CCD equations into these three channels but which are not present in the typical RPA approach.  This choice turns out not to be associated with a symplectic RPA-like matrix.  Rather, we must include the interaction $v_{ij}^{ab}$ with a positive sign and the remainder of the driver term (the terms giving rise to orbital energy denominators, in other words) we can safely leave with a negative sign.  This gives us
\begin{equation}
v_{ij}^{ab} - F^a_c \, t_{ij}^{cb} - F^b_c \, t_{ij}^{ac} + F^k_i \, t_{kj}^{ab} + F_j^k \, t_{ik}^{ab} - v_{ic}^{kb} \, t_{kj}^{ac} - v_{jc}^{ka} \, t_{ik}^{cb} - t_{ik}^{cb} \, v^{kl}_{cd} \, t_{lj}^{ad} = 0.
\end{equation}
This can be fruitfully rewritten as
\begin{equation}
-v_{ij}^{ab} + \left(F^a_c \, \delta^k_j - F^k_j \, \delta^a_c + v_{jc}^{ka}\right) t_{ik}^{cb} + \left(F^b_c \, \delta^k_i - F^k_i \, \delta^b_c + v_{ic}^{kb}\right) t_{kj}^{ac} -  t_{ik}^{cb} \, v^{kl}_{dc} \, t_{lj}^{ad} = 0.
\end{equation}
This time, we want to define composite indices as $ib$ and $ja$; this prevents us from simply adding the ring and crossed-ring equations.  We can define
\begin{subequations}
\begin{align}
\tilde{A}_{kc,ja} &= F^a_c \, \delta^k_j - F^k_j \, \delta^a_c + v_{jc}^{ka}
\\
\tilde{B}_{kc,ld} &= v^{kl}_{dc}
\end{align}
\end{subequations}
where $\tilde{\bm{A}}$ and $\tilde{\bm{B}}$ are closely related to the $\bm{A}$ and $\bm{B}$ matrices of particle-hole RPA, but differ by the sign of the two-electron integral.  In terms of these newest quantities, the crossed-ring--CCD looks like
\begin{equation}
\bm{0} = -\tilde{\bm{B}}^\star + \tilde{\bm{T}} \, \tilde{\bm{A}} + \tilde{\bm{A}}^\star \, \tilde{\bm{T}} - \tilde{\bm{T}} \, \tilde{\bm{B}} \, \tilde{\bm{T}}.
\end{equation}
From our discussion of particle-hole RPA, it should be clear that this is the Riccati equation corresponding to the symplectic eigenvalue problem
\begin{equation}
\begin{pmatrix} \tilde{\bm{A}} & -\tilde{\bm{B}} \\ \tilde{\bm{B}}^\star & -\tilde{\bm{A}}^\star \end{pmatrix}
\begin{pmatrix} \bm{X} \\ \bm{Y} \end{pmatrix}
=
\begin{pmatrix} \bm{X} \\ \bm{Y} \end{pmatrix}
\bm{\varpi}.
\end{equation}

\subsection{Combining the Three Channels}
Returning to the Brueckner CCD amplitude equations, we see that we can write them
\begin{equation}
0 = v + \left(\mathrm{Ring} - v\right) + \left(\mathrm{Ladder} - v\right) + \left(\textrm{Crossed-Ring} - v\right).
\end{equation}
We emphasize that we cannot simply add the Riccati equations for the three channels, as the bosonic composite external indices are formed from the fermionic indices in three different ways.  We also emphasize that including the mosaic terms in defining an effective one-body Hamiltonian is necessary to leave this simple form.  Finally, note that the three Riccati equations, corresponding to three different RPA-like problems, are tied together by the requirement that the amplitudes $t$ are the same -- these amplitudes, in other words, force the channels to interact.  This is made possible by the inclusion of the crossed-ring terms, which provide the necessary antisymmetrization that the ring channel lacks.

The patching of these three problems as a conceptual tool for understanding CCD has never been discussed in the literature, to the best of our knowledge.  From the analysis above, we see that the CC equations can be interpreted as the sum of three quadratic bosonic problems: rings, ladders, and crossed-rings with a renormalized one-body Hamiltonian ($F$) and the regular Coulomb two-body interaction. Each of the channels contract the CC amplitudes in a different manner. In this sense, we could say that \textit{CC theory is as a correct bosonization} of fermion excitations because it yields fully antisymmetrized excitation amplitudes fulfilling Pauli's principle, together with a well-defined wavefunction and corresponding density matrices.

\section{Results}
In this section, we provide a few numerical results, showing the relative importance of the particle-particle and particle-hole channels of RPA, as well as the importance of the various contributions to the CCD equations.  Variants of CCD will be identified by whether they include ladder terms (``l''), ring terms (``r''), and mosaic terms (``m''); thus, ring-CCD in this notation is rCCD and ladder-CCD is l-CCD.  Because they do not appear in RPA, we do not include crossed-ring terms in this section except in the form of the full CCD.  We shall see later that while the crossed-rings are vital for restoring the full fermionic character of the CCD wavefunction, they can also unbalance the CC equations when other classes of diagrams are omitted.

All results were generated using an in-house program.  As this paper is not intended to generate benchmark data or even comparisons with experiment, but rather to discuss qualitative features, we will use small basis sets and not worry about basis set incompleteness error.

We begin our discussion with the dissociation of H$_2$, which is of course paradigmatic in quantum chemistry.  As is well known, the restricted Hartree-Fock (RHF) solution goes to much too high an energy, as a result of contamination from ionic dissociation fragments.  This is remedied (energetically) by unrestricted Hartree-Fock (UHF), which dissociates correctly at the cost of broken spatial and spin symmetries.  There is thus an instability from RHF to UHF past the Coulson-Fischer point, and beyond this point particle-hole RPA based on the RHF reference yields unphysical complex correlation energies.

\begin{figure}
\includegraphics[width=0.65\textwidth]{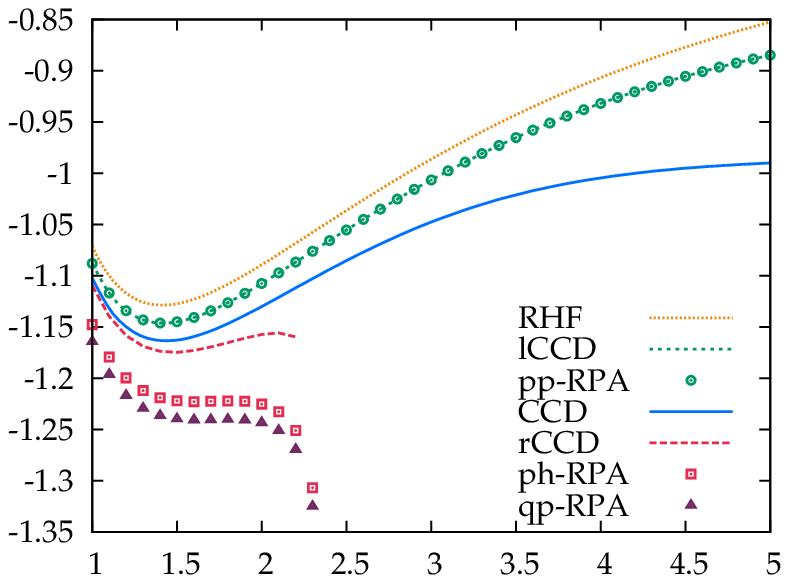}
\\
\includegraphics[width=0.65\textwidth]{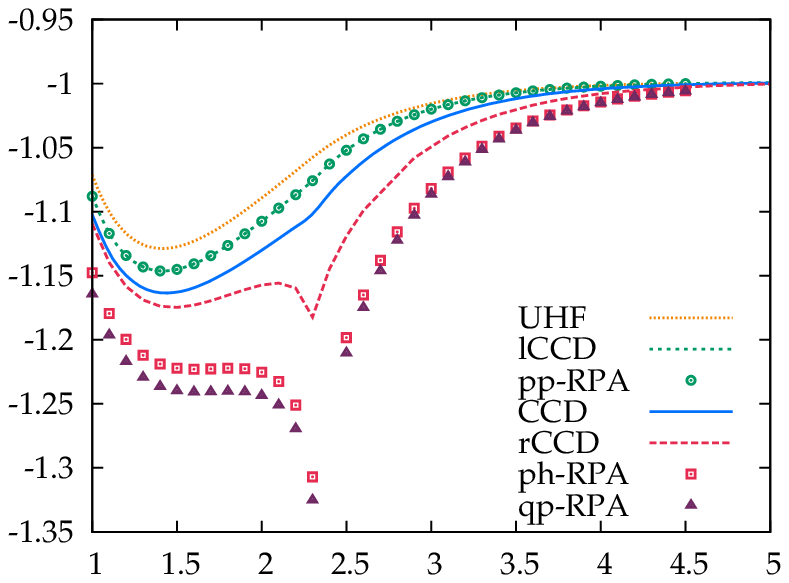}
\caption{Dissociation curve of H$_2$ in the cc-pvdz basis set, obtained with various flavors of RPA, compared to CCD, ring-CCD, and ladder-CCD.  Top panel: RHF reference.  Bottom panel: UHF reference.
\label{Fig:H2RPA}}
\end{figure}

The RPA dissociation of H$_2$ is shown in Fig. \ref{Fig:H2RPA}.  There are several key features here we wish to point out.  First, the ladder-CCD and the particle-particle RPA energies are indeed identical, bearing out our anayltic proof earlier.  While particle-particle RPA undercorrelates, particle-hole RPA overcorrelates.  Since quasiparticle RPA includes both, it overcorrelates even more.  Near the Coulson-Fischer point, particle-hole RPA is particularly bad, with a cusp at the Coulson-Fischer point; this behavior is inherited by quasiparticle RPA.  If one uses an RHF reference instead of a UHF reference, the particle-hole and quasiparticle RPA energies become complex.  Though the particle-particle RPA remains well behaved, its undercorrelation is greatly exaggerated.

\begin{figure}
\includegraphics[width=0.65\textwidth]{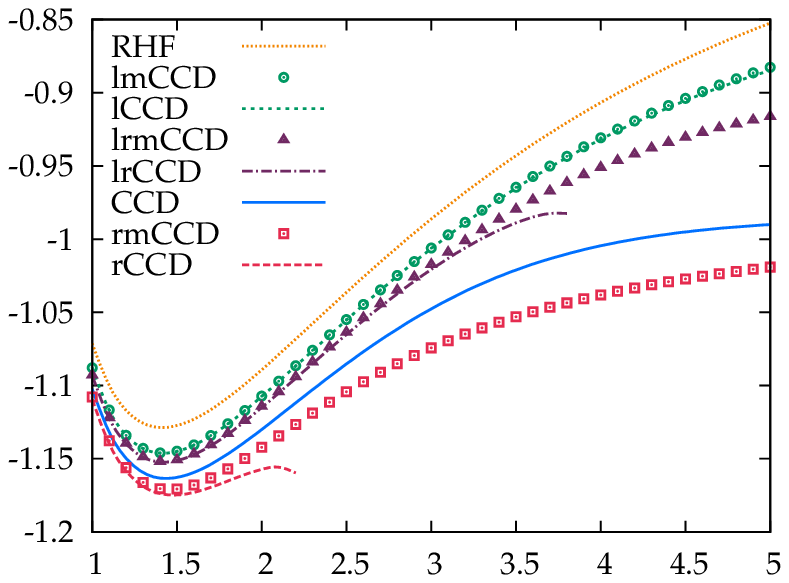}
\\
\includegraphics[width=0.65\textwidth]{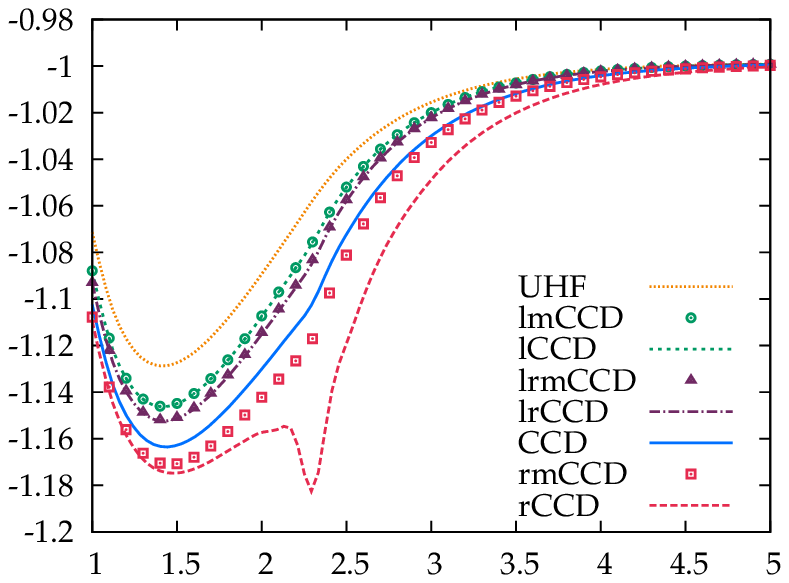}
\caption{Dissociation curve of H$_2$ in the cc-pvdz basis set, obtained with variants of CCD.  Top panel: RHF reference.  Bottom panel: UHF reference.
\label{Fig:H2CCD}}
\end{figure}

In Fig. \ref{Fig:H2CCD}, we show results for variants of CCD including selective terms.  Let us start with results based on the RHF reference.  In this case, as is well-known, the ring-CCD does not converge past the Coulson-Fischer point.  Our results are nevertheless sufficient to illustrate that ladder-CCD undercorrelates while ring-CCD overcorrelates.  Including both ladders and rings undercorrelates, but for a sufficiently stretched bond, we were unable to converge the equations.  Adding the mosaic terms has, in general, a relatively small effect, but note that they cure the convergence difficulties we face due to the inclusion of the ring diagrams without corresponding inclusion of the crossed rings.

The story is qualitatively similar using the UHF reference.  Again, the ladder-CCD undercorrelates while the ring-CCD overcorrelates.  Including both ladders and rings qualitatively resembles ladder-CCD.  In this case, the ring-CCD equations can be converged to dissociation, but the curve near the Coulson-Fischer point remains pathological.  Again, the mosaic terms have a relatively small effect for the most part, but cure the worst of the pathologies of ring-CCD.  We note finally that CCD itself has a shoulder at the Coulson-Fischer point, which would be cured by the inclusion of single excitations.  Equivalently, we could iterate the BD equations, in which case the mosaic terms would be naturally taken care of and the CCD would become exact.

\begin{figure}
\includegraphics[width=0.65\textwidth]{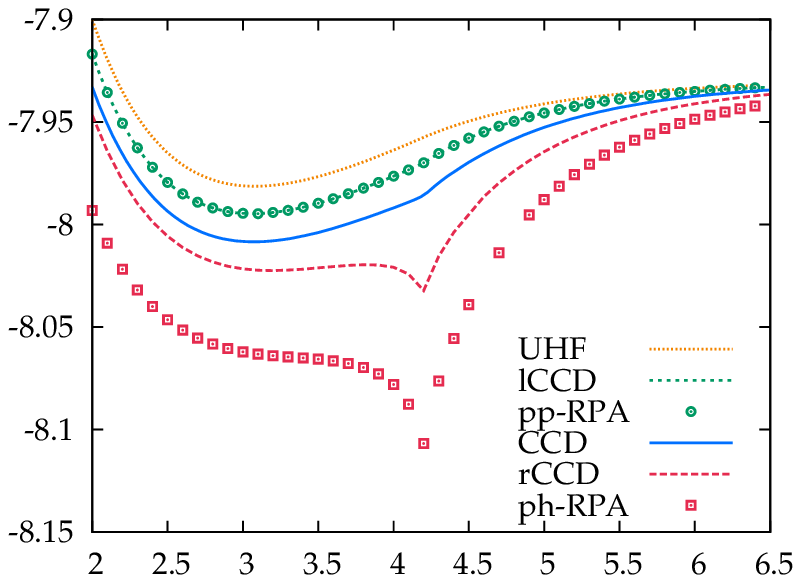}
\caption{Dissocation curve of LiH in the 6-31G** basis set, obtained with variants of RPA
\label{Fig:LiHRPA}}
\end{figure}

\begin{figure}
\includegraphics[width=0.65\textwidth]{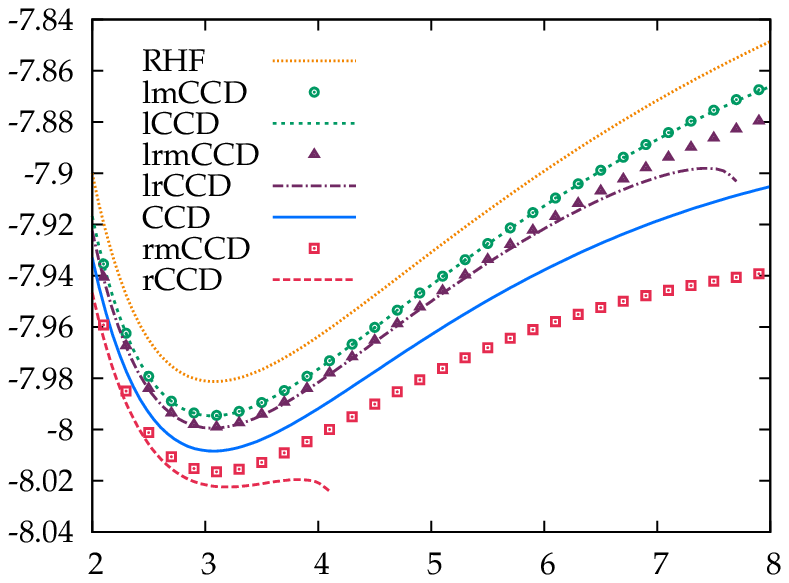}
\\
\includegraphics[width=0.65\textwidth]{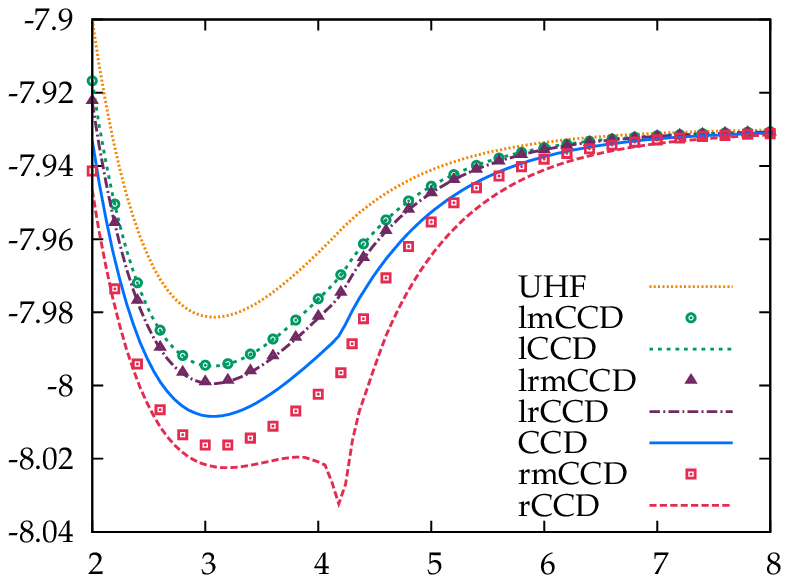}
\caption{Dissociation curve of LiH in the 6-31G** basis set, obtained with variants of CCD.  Top panel: RHF reference.  Bottom panel: UHF reference.
\label{Fig:LiHCCD}}
\end{figure}

We next turn to the dissociation of LiH, for which we use the 6-31G** basis set.  In Fig. \ref{Fig:LiHRPA} we show UHF-based RPA dissociation curves, while Fig. \ref{Fig:LiHCCD} shows both RHF- and UHF-based CCD dissociations.  The main details remain the same as in H$_2$.  Ring-CCD overcorrelates significantly, and particle-hole accordingly overcorrelates even more badly.  In contrast, ladder-CCD = particle-particle RPA undercorrelates.  Including both ladders and rings in quasiparticle RPA overcorrelates even more badly than does particle-hole RPA (not shown), while including them both in a coupled cluster approach improves but does not fully cure the undercorrelation from ladder-CCD.  Starting from the RHF reference, ring-CCD does not converge past the Coulson-Fischer point and lr-CCD eventually stops converging as well.  These convergence problems are cured by including the mosaic terms.  Starting from the UHF reference, ring-CCD (and therefore particle-hole RPA) exhibits pathological behavior in the vicinity of the Coulson-Fischer point, which is again cured by adding the mosaic terms.  In all other cases, the mosaic terms have essentially negligible effects.

\begin{figure}
\includegraphics[width=0.65\textwidth]{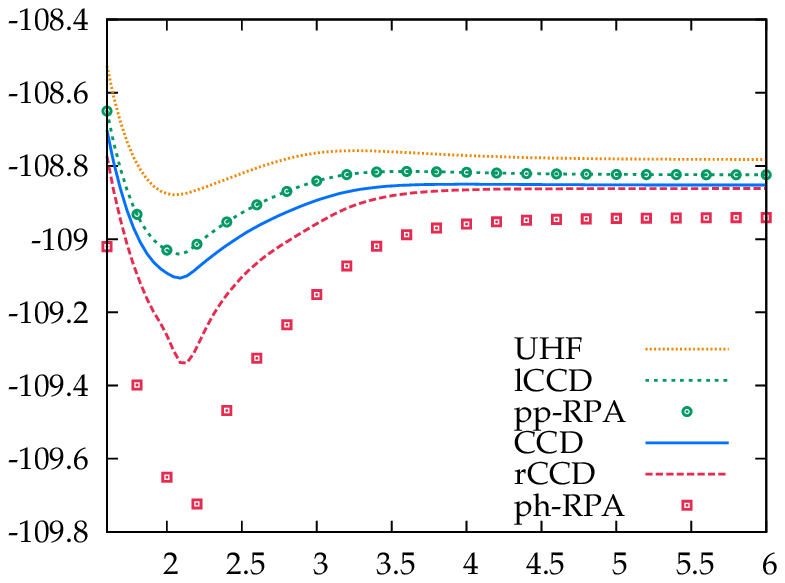}
\\
\includegraphics[width=0.65\textwidth]{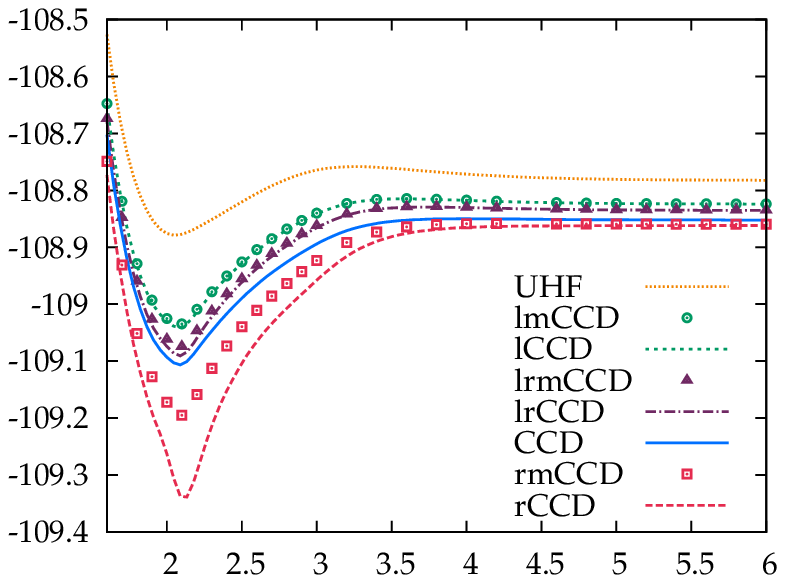}
\caption{Dissociation curve of N$_2$ in the cc-pvdz basis set with $d$ functions removed.  Top panel: RPA variants and related CCD approximations.  Bottom panel: CCD variants including mosaic terms.
\label{Fig:N2}}
\end{figure}

Lastly, we consider the dissociation of N$_2$, which we have examined in the cc-pVDZ basis set with $d$ functions removed for ease of convergence, and since we remain interested only in the qualitative picture.  Results are shown in Fig. \ref{Fig:N2}, using a broken-symmetry reference.  Yet again, the particle-hole RPA overcorrelates, as does the ring-CCD, while the particle-particle RPA undercorrelates.  As usual, the particle-hole RPA and ring-CCD display especially severe problems near the Coulson-Fischer point.  The mosaic terms are again fairly unimportant, except in the case of ring-CCD where they offer a large improvement.  Including both ladders and rings in the CCD is superior to including just one or the other.

\section{Discussion}
It should be clear from previous sections that there is a recurrying theme in all of these approaches, which is the attempt to treat the ground state with a fermionic wavefunction which is to be correlated via \textit{bosonic} excitations.  This seems like a natural way of using fermion and boson mean-field methods.  The main difficulty is that composite fermions are not actually bosonic, so whatever sort of bosonization one undertakes can only be approximate when the composite indices are never broken (as in the various flavors of RPA).  A secondary difficulty is that bosonizing the excitations, which are $\mathcal{O}(M^2)$ in number where $M$ is the number of basis functions, leads to an $\mathcal{O}(M^6)$ bosonic mean-field treatment, so that the computational scaling does not improve upon traditional CCD without further approximation.  This is true both for particle-hole and for particle-particle RPA.  Note that in practice, the scaling of RPA can often be considerably reduced through techniques such as Cholesky decomposition and variants.\cite{RPAReview2}

The traditional CCD equations achieve a sort of correct bosonization by including not only the particle-particle (ladder) and particle-hole (ring) channels, but adding crossed-ring terms as well.  Insisting, additionally, that the amplitudes obtained from the corresponding symplectic eigenvalue problems all be identical once the composite fermion indices are interpreted forces the three channels to interact, and in this way overcomes the overcorrelation of quasiparticle RPA while correctly enforcing fermionic antisymmetry.

\begin{figure}
\includegraphics[width=0.65\textwidth]{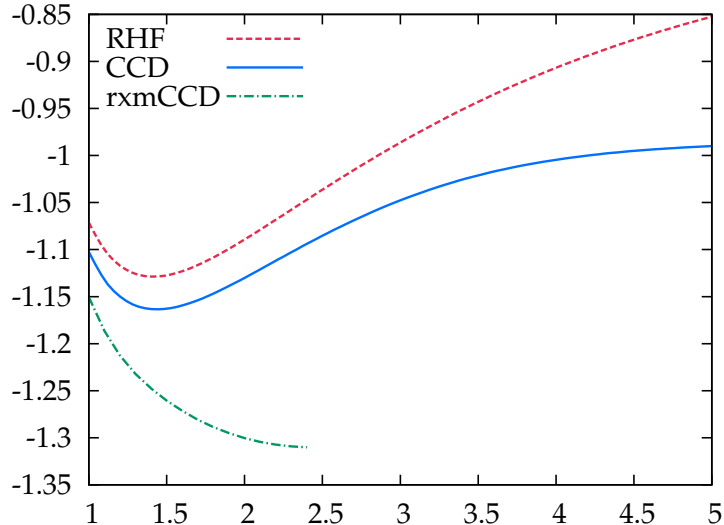}
\caption{RHF-based dissociation curves for H$_2$ in the cc-pVDZ basis set.  We include RHF, CCD, and a variant of CCD which exclude the latter diagrams (rxmCCD).  Beyond the Coulson-Fischer point, the latter does not converge.
\label{Fig:H2rxm}}
\end{figure}

From one perspective, the main function of the crossed-ring terms in the CCD equations is simply to guarantee this fermionic antisymmetry.  Including the crossed-ring terms is not, however, the only way to impose this constraint.  Second-order screened exchange\cite{SOSEX} (SOSEX) does this for the case of ring-CCD.  In SOSEX, one first solves the ring-CCD equations, then antisymmetrizes the resulting amplitudes before calculating the correlation energy (using, it should be noted, the correct coupled cluster factor of $\tfrac{1}{4}$).  This reduces the ring-CCD correlation energy substantially, preventing the dramatic overcorrelation which we have seen.  One might think that a better approach might be to include both rings and crossed-ring in a coupled cluster approach.  This, however, is unsuccessful.  Including only rings and crossed-rings in a CCD framework, we encounter severe convergence difficulties; when we can converge the results, they are quite poor.  Adding the mosaic terms as well alleviates these convergence difficulties, but overcorrelates terribly, as seen in Fig. \ref{Fig:H2rxm}.

While the crossed-ring terms can be naturally incorporated into the CCD framework, their origin in RPA of all flavors is obscure.  In fact, they seem entirely artificial from the RPA perspective.  A proper RPA-style mechanism by which these antisymmetrizing pieces can be included would most likely be quite valuable.

An important point to bear in mind in quasiparticle RPA is that the 13 and 31 blocks of the Hamiltonian do not contribute at all.  We can imagine a Brueckner-style renormalization such as
\begin{equation}
\tilde{H}_{pq}^{11} = H_{pq}^{11} + \left[H^{31}_{prst} \left(\bm{\mathcal{Y}} \, \bm{\mathcal{X}}^{-1}\right)_{stqr} + \textrm{h.c.}\right]
\end{equation}
where, in analogy with the Brueckner effective Hamiltonian, the effects of $H^{31}$ and $H^{13}$ are incorporated iteratively by changing the reference quasiparticle determinants orbitals and energies.  One might hope that the dressed Fock and pairing fields caused by this Brueckner-style renormalization would induce number fluctuations in the reference determinant.  Were this to happen, the particle-particle and particle-hole channels would interact.  This might cure some of the overcorrelation endemic to quasiparticle RPA.  Recall that for a standard repulsive two-body interaction, the HFB method is entirely equivalent to Hartree-Fock.

Overall, we recommend a careful combination of particle-particle and particle-hole RPA if one wishes to work within an RPA framework.  The two methods are in some sense complementary, and each describes different physics.  Particle-hole RPA, for example, works quite well for long-range electron-electron interactions, and successfully incorporates van der Waals binding.\cite{Dobson}  On the other hand, from the form of the bosonic excitation operators, we see that particle-particle RPA should be suitable for the description of charge fluctuations which are beyond the scope of conventional particle-hole RPA.  These recommendations are particularly important in efforts to incorporate RPA correlation effects into density functional theory.\footnote{While this paper was about to be submitted, we learned about an effort to incorporate particle-particle RPA correlation in DFT.  See H. van Aggelen, Y. Yang, and W. Yang, (2013) arXiv:1306.4957, and D. Peng, S. N. Steinmann, H. van Aggelen, and W. Yang, (2013) arXiv:1306.5638}  We have seen, moreover, that the CCD result often lies between ladder-CCD and ring-CCD, and thus between particle-hole and particle-particle RPA.  Finally, what we have christened the mosaic terms are also apparently quite important, and should perhaps be included in post--Hartree-Fock RPA schemes.

\section{Acknowledgments}
This work was supported by the National Science Foundation CHE-1110884 and the Welch Foundation (C-0036).

\bibliography{ppRPA}

\end{document}